\documentclass[aps,prx,superscriptaddress,amsmath,amssymb,twocolumn,showpacs,floatfix,reprint]{revtex4-2}
\usepackage[colorlinks=true, urlcolor=blue, linkcolor=blue, citecolor=blue, pdftex]{hyperref}
\usepackage[capitalise]{cleveref}
\usepackage{xr}
\usepackage{multirow}
\usepackage[dvipsnames]{xcolor}
\usepackage[utf8]{inputenc}
\usepackage{braket}
\usepackage{graphicx}
\usepackage{makecell}
\usepackage{balance}
\usepackage{changes}
\usepackage{comment}

\begin{document}
\title{Beyond Variational Bias: Resolving Intertwined Orders in the Hubbard Model}

\author{Luciano Loris Viteritti}
\thanks{These authors contributed equally. Correspondence should be addressed to rrende@flatironinstitute.org and luciano.viteritti@epfl.ch}
\affiliation{Institute of Physics, \'{E}cole Polytechnique F\'{e}d\'{e}rale de Lausanne (EPFL), CH-1015 Lausanne, Switzerland}

\author{Riccardo Rende}
\thanks{These authors contributed equally. Correspondence should be addressed to rrende@flatironinstitute.org and luciano.viteritti@epfl.ch}
\affiliation{Center for Computational Quantum Physics, Flatiron Institute, 162 5th Avenue, New York, NY 10010}

\author{Christopher Roth}
\affiliation{Center for Computational Quantum Physics, Flatiron Institute, 162 5th Avenue, New York, NY 10010}

\author{Anirvan Sengupta}
\affiliation{Center for Computational Quantum Physics, Flatiron Institute, 162 5th Avenue, New York, NY 10010}
\affiliation{Center for Computational Mathematics, Flatiron Institute, 162 5th Avenue, New York, NY 10010}
\affiliation{Department of Physics and Astronomy, Rutgers University, Piscataway, New Jersey 08854, USA}

\author{Giuseppe Carleo}
\affiliation{Institute of Physics, \'{E}cole Polytechnique F\'{e}d\'{e}rale de Lausanne (EPFL), CH-1015 Lausanne, Switzerland}

\author{Antoine Georges}
\affiliation{Center for Computational Quantum Physics, Flatiron Institute, 162 5th Avenue, New York, NY 10010}
\affiliation{Coll{\`e}ge de France, 11 place Marcelin Berthelot, 75005 Paris, France}
\affiliation{CPHT, CNRS, {\'E}cole Polytechnique, IP Paris, F-91128 Palaiseau, France}
\affiliation{DQMP, Universit{\'e} de Gen{\`e}ve, 24 quai Ernest Ansermet, CH-1211 Gen{\`e}ve, Suisse}

\date{\today}

\begin{abstract}
The two-dimensional Hubbard model at finite doping hosts competing or intertwined orders, resulting in conflicting conclusions from different computational approaches regarding its ground state. We show that a key source of such discrepancies is the bias encoded in the variational {\it ansatz}. We consider three different Transformer backflow fermionic wave functions based on a Slater determinant, its particle–hole counterpart, and a Pfaffian, initialized without any mean-field pretraining. We show that, despite achieving nearly degenerate, state-of-the-art variational energies, each {\it ansatz} converges to a state with qualitatively different spin, charge, and pairing correlations. Upon improving accuracy via symmetry restoration and variance reduction, however, all three converge to the same physical picture: coexisting superconducting and stripe orders. These results demonstrate that variational energy alone is insufficient to identify the ground state in the presence of competing phases, and highlight the importance of tracking how correlation functions evolve as the wave function is systematically improved before drawing physical conclusions.
\end{abstract}

\maketitle
\section{Introduction} 
The two-dimensional Hubbard model occupies a central position in the theory of correlated electrons~\cite{hubbard1963, white1989, gull2015, xu2024, simkovic2024}. Although it is one of the simplest models for the description of strongly correlated fermions, it exhibits a remarkably rich phenomenology upon doping~\cite{shiwei2020, sorella2023}. In the context of high-$T_c$ cuprate superconductivity, it is widely used as a minimal setting in which antiferromagnetism, charge modulations, and unconventional pairing can emerge from the same microscopic ingredients. Despite decades of work, however, the nature of the dominant correlations in the doped regime, and their evolution with interaction strength, next-nearest-neighbor hopping, and carrier density, remain actively debated~\cite{gull2015, xu2024}. An explicit treatment of electron–electron interactions leads to an exponential growth of the Hilbert space with system size, making exact solutions intractable in general. Accurate approximate methods are therefore essential to access the ground-state properties. In this context, variational wave-function approaches have played an essential role, ranging from physically motivated Jastrow–Slater states~\cite{capello2005, tocchio2008, ido2018, sorella2023} and tensor-network-related methods based on Density Matrix Renormalization Group (DMRG)~\cite{hager2005, zheng2017, ehlers2017, jiang2019, scheb2023, liu2025, jiang2025} to, more recently, Neural-Network Quantum States (NQS)~\cite{nomura2017, luo2019, robledo2022, gu2025,Chen2025NNPfaffian, Roth2025SCNQS, sharma2025, ibarra2025}.

%%%%%%%%%%%%%%%%%%%%%%%%%%%%%%%%
\begin{figure}[t]
    \begin{center}
\centerline{\includegraphics[width=\columnwidth]{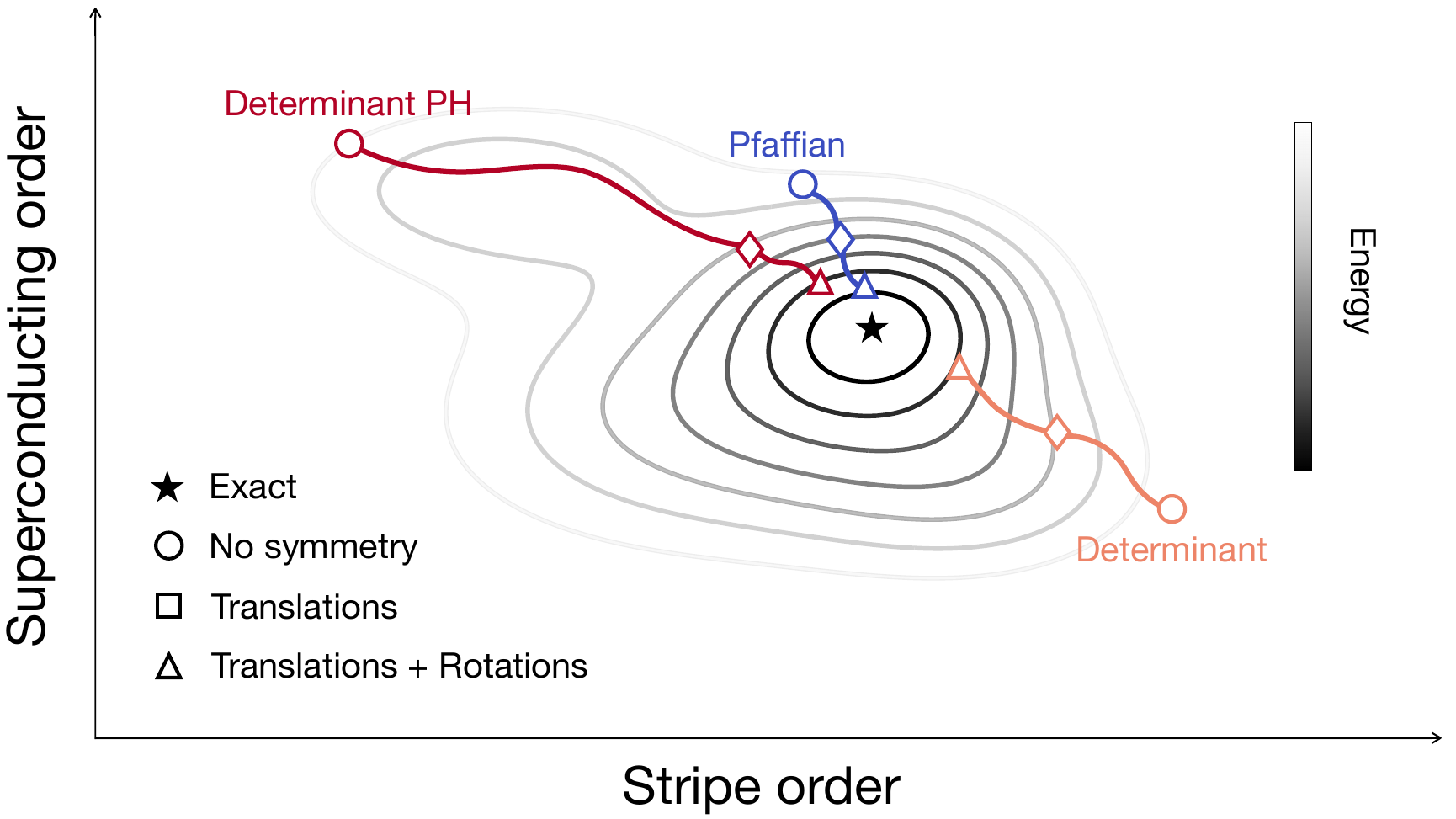}}
        \caption{\label{fig:pictorial} Schematic illustration of the optimization landscape for different antisymmetrization schemes. Contours represent variational energy levels. Starting from random initialization and without symmetry constraints, determinant, determinant–particle-hole, and Pfaffian constructions converge toward qualitatively distinct solutions with nearly degenerate energies but different dominant orders. Upon variational energy reduction and symmetry restoration, all trajectories approach the same correlated ground state in which superconducting and stripe order coexist.}
    \end{center}
\end{figure}
%%%%%%%%%%%%%%%%%%%%%%%%%%%%%%%%

A central challenge in determining the ground-state properties of interacting fermionic systems is the possible close energetic competition among phases with distinct long range order, which renders crucial the choice of the variational {\it ansatz}. The two-dimensional $t$-$t'$ Hubbard model provides a paradigmatic example: distinct variational constructions can converge to nearly degenerate energies while supporting qualitatively different physical states~\cite{gu2025,Roth2025SCNQS,shiwei2020,jiang2025}. Variational approaches inevitably introduce biases through architectural constraints, such as the underlying graph and finite bond dimension in tensor networks~\cite{jiang2025, liu2025}, fixed unit cells in iPEPS~\cite{corboz2010} or specific design choices in NQS architectures. 
These constraints restrict the class of efficiently representable wave functions and shape the optimization landscape, potentially favoring certain orders over others. A systematic understanding and control of these biases is, therefore, essential for constructing reliable variational descriptions of correlated fermionic states.

In this work, we address this fundamental problem using neural-network backflow 
wave functions~\cite{diluo_prl, prl_backflow_holzmann} with distinct architectural biases, taking the two-dimensional $t$–$t^{\prime}$ Hubbard model at finite doping as a representative model.
Employing a Transformer-based neural-network architecture~\cite{gu2025,sharma2025,ma2025,viteritti2023prl,viteritti2025prb,rende2026alm, rende2025} to parametrize the backflow orbitals, we demonstrate that distinct antisymmetrization schemes, specifically a Slater determinant in the standard and particle–hole (PH) representations and a Pfaffian wave function, converge to physically distinct variational states when optimized from random initializations, in the absence of mean-field pretraining or auxiliary symmetry-breaking fields such as pinning potentials~\cite{sorella2023, gu2025}. The resulting ground-state candidates exhibit competing orders, ranging from stripe-dominated phases stabilized by the determinant backflow to strongly superconducting states favored by the determinant in the particle-hole representation, despite exhibiting nearly degenerate, variational energies. 
This observation makes it clear that variational energy alone is an unreliable proxy for assessing whether a variational wavefunction captures the correct ground state physics.

Crucially, this issue can be resolved by restoring symmetries. After adding translational and point group symmetries~\cite{loehr2025}, all ans\"atze converge to a physical state characterized by consistent charge, spin, and pairing correlations. This convergence resolves the ambiguity and establishes a robust, representation-independent characterization of the ground-state physical properties. A pictorial illustration of this behavior is shown in \cref{fig:pictorial}. 

The central message of this work is that, in challenging models such as the doped Hubbard case, where several orders compete or coexist, the chosen variational representation can introduce a strong architectural bias and thereby alter the physical picture. This makes it essential to control and systematically improve 
the expressivity and accuracy of the variational state, in particular by reducing its variance in an unbiased manner, before drawing conclusions about the nature of the true ground state. In this context, we find that the antisymmetrization scheme induces a bias on the variational wavefunction. For each of the three cases, the NQS is biased towards the phases that it can represent at the mean field level, i.e., when the orbitals are set to be configuration independent. Hence, the determinant tends to overestimate stripe order and underestimate pairing, this is reversed under particle-hole transformation. The Pfaffian, which can 
simultaneously 
represent these coexisting orders at the mean field level~\cite{Roth2025SCNQS}, tends to produce the most accurate correlation functions before symmetrization, although it still overestimates pairing. Crucially, by tracking the evolution of these correlation functions under variance reduction, this bias can be revealed, and the ground-state physics can be established more reliably.

Finally, our results provide accurate ground state properties for the $8\times 8$ Hubbard model on a square lattice with periodic boundary conditions, at $t'=-0.2$, $U=8$, and $1/8$ hole doping. We propose this setup as a challenging benchmark for numerical methods targeting fermionic systems.

%%%%%%%%%%%%%%%%%%%%%%%%%%%%%%%%
\begin{figure*}[t]
    \begin{center}
\centerline{\includegraphics[width=2.0\columnwidth]{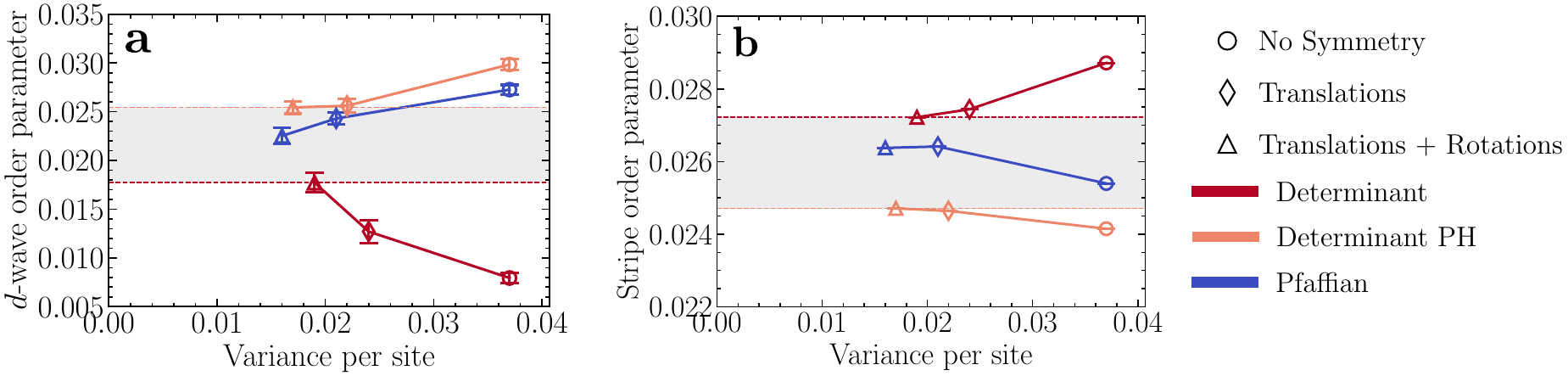}}
        \caption{\label{fig:order_par}  Superconducting and stripe order parameters as a function of variational accuracy for the $8\times8$ Hubbard model under PBC at $U=8.0$, $t^{\prime}=-0.2$, and $\delta=1/8$ [see \cref{eq:Hubbard}]. \textbf{Panel a:} $d$-wave pairing order parameter $\Delta_{\text{SC}}$ versus energy variance per site. \textbf{Panel b:} Stripe spin order parameter $M^2_{\text{str}}$ versus variance per site. The definitions of the order parameters are given in the main text. In panels (a) and (b), different colors correspond to the three antisymmetrization schemes: determinant (red), determinant in particle hole (PH) (blue) and Pfaffian (orange). Different marker symbols indicate the degree of symmetry restoration without symmetry projections: no symmetry projection (circles), translational symmetry restored (rhombi), and both translational and rotational symmetries restored (triangles). Dashed horizontal lines mark the values obtained from the fully symmetrized determinant and particle-hole determinant states. These are interpreted as lower and upper bounds for the exact value of the corresponding order parameter, while the shaded region indicates the interval in which the exact value is expected to lie. See \cref{sec:bounds} of the \textit{Appendix} for further details on the definition of the bounds.}
    \end{center}
\end{figure*}
%%%%%%%%%%%%%%%%%%%%%%%%%%%%%%%%

\section{Two dimensional $t$–$t’$ Hubbard model}
The $t$–$t’$ Hubbard model is defined by the following Hamiltonian
\begin{equation}
\hat H =
- \sum_{\boldsymbol{r},\boldsymbol{r}^{\prime},\sigma} t_{\boldsymbol{r}\boldsymbol{r}^{\prime}}
(\hat c^\dagger_{\boldsymbol{r}\sigma} \hat c_{\boldsymbol{r}^{\prime}\sigma} + \text{h.c.})
+ U \sum_{\boldsymbol{r}} \hat n_{\boldsymbol{r}\uparrow} \hat n_{\boldsymbol{r}\downarrow} \ ,
\label{eq:Hubbard}
\end{equation}
where $\boldsymbol{r}$,$\boldsymbol{r}^{\prime}$ indicate the sites of a $L\times L$ square lattice with periodic boundary conditions. Here, $\hat c^\dagger_{\boldsymbol{r}\sigma}$ ($\hat c_{\boldsymbol{r}\sigma}$) creates (annihilates) a fermion with spin $\sigma \in \{\uparrow,\downarrow\}$ on site $\boldsymbol{r}$, and $\hat n_{\boldsymbol{r}\sigma} = \hat c^\dagger_{\boldsymbol{r}\sigma} \hat c_{\boldsymbol{r}\sigma}$ is the particle number operator. The hopping amplitudes $t_{\boldsymbol r \boldsymbol r’}$ account for both nearest-neighbor hopping, $t_{\boldsymbol r \boldsymbol r’} = t$, and next-nearest-neighbor hopping, $t_{\boldsymbol r \boldsymbol r’} = t’$, whereas $U$ denotes the on-site Coulomb repulsion. In the following, we set the hopping amplitude to $t = 1$ and express all remaining couplings in units of $t$. The filling is controlled by fixing the particle number $N_e$ and consequently the hole doping is $\delta=1-n$, where $n=N_e/N$ is the particle density with $N=L^2$ the total number of sites. 

In addition to the standard fermionic representation, we also consider a particle-hole (PH) formulation. This is done by rewriting the hamiltonian in terms of transformed fermionic operators  ${\hat c_{\boldsymbol{r}\uparrow} \rightarrow \hat c_{\boldsymbol{r}\uparrow}}$ and ${\hat c_{\boldsymbol{r}\downarrow} \rightarrow \,\hat c_{\boldsymbol{r}\downarrow}^\dagger}$.
The advantage of this representation is that mean-field superconducting states, such as the ground state of the Bardeen–Cooper–Schrieffer Hamiltonian~\cite{bardeen1957}, can be expressed exactly as a determinant, providing a useful point of comparison when analyzing superconducting correlations (see \cref{sec:antisymm_schemes} of the \textit{Appendix}).

\section{Characterization of Pairing and Stripe order}\label{sec:results}
\Cref{fig:order_par} provides a quantitative overview of our results, showing the superconducting and spin order parameters as functions of the variational accuracy. The calculations are performed using the Transformer backflow {\it ansatz}~\cite{gu2025, sharma2025, rende2026alm} (see \cref{sec:transformer} of the \textit{Appendix} for additional details) with different antisymmetrization output layers that encode distinct biases, namely a Slater determinant, a particle–hole transformed determinant, and a Pfaffian. We focus on an $8\times 8$ cluster with $\delta=1/8$, $t^{\prime}=-0.2$, and $U=8.0$, a standard parameter set representative of the superconducting regime of the Hubbard model.

Panel (a) of \cref{fig:order_par} shows the $d$-wave pairing order parameter $\Delta_{\text{SC}}$, extracted from the long-distance behavior of the pair–pair correlation function. We define ${\Delta_{\text{SC}}= \sqrt{\tfrac{1}{\mathcal{M}}\sum_{|\boldsymbol{r}|\ge r_{\text{max}}} C_p(\boldsymbol{r})}}$ where $\mathcal{M}$ is the number of vectors satisfying $|\boldsymbol{r}|\ge r_{\text{max}}$. For $L=8$ we set $r_{\text{max}}=4$. The pairing correlation function is~\cite{Roth2025SCNQS}:
\begin{equation}\label{eq:pairing_corr}
    C_p(\boldsymbol{r}) = \langle\hat\Delta_{\boldsymbol{0}}^{\dagger}\hat\Delta_{\boldsymbol{r} }\rangle - \mathcal{N}_{\boldsymbol{0}, \boldsymbol{r}} \ ,
\end{equation}
where $\hat\Delta_{\boldsymbol{r}}=\frac{1}{4}\sum_{\boldsymbol{\eta}} h_{\boldsymbol{r}, \boldsymbol{\eta}}\ \hat{c}_{\boldsymbol{r},\uparrow}\hat{c}_{\boldsymbol{r}+\boldsymbol{\eta},\downarrow}$.
Here the sum over $\boldsymbol\eta$ runs over the nearest neighbors of site $\boldsymbol{r}$, and the structure of $h_{\boldsymbol{r},\boldsymbol\eta}$ select the $d_{x^2-y^2}$ pairing order: $h_{\boldsymbol{r},\boldsymbol\eta}=+1$ for horizontal neighbors and $h_{\boldsymbol{r},\boldsymbol\eta}=-1$ for vertical neighbors~\cite{shiwei2020}. We subtract the disconnected (normal-state) contribution, keeping only the term which conserve the number of particles and total $S^z$~\cite{Chen2025NNPfaffian}:
\begin{equation}
    \mathcal{N}_{\boldsymbol{0}, \boldsymbol{r}}=\frac{1}{16}\langle \hat{c}^\dagger_{\boldsymbol{r},\uparrow}\hat{c}_{\boldsymbol{0},\uparrow}\rangle\sum_{\boldsymbol{\eta},\boldsymbol{\eta}^{\prime}}h_{\boldsymbol{0}, \boldsymbol{\eta}}h_{\boldsymbol{r}, \boldsymbol{\eta}^{\prime}}\langle \hat{c}^\dagger_{\boldsymbol{r}+\boldsymbol{\eta}^{\prime},\downarrow}\hat{c}_{\boldsymbol{0}+\boldsymbol{\eta},\downarrow}\rangle \ .
\end{equation}
Panel (b) of \cref{fig:order_par} displays the stripe spin order parameter, extracted from the peaks of the spin structure factor  ${S(\boldsymbol{k})=\sum_{\boldsymbol{r}}e^{i\boldsymbol{k}\cdot \boldsymbol{r}}}C_S(\boldsymbol{r})$, where ${C_S(\boldsymbol{r})=\braket{\hat{{S}}^z_{\boldsymbol{0}}\hat{{S}}^z_{\boldsymbol{r}}}}$. To allow for a direct comparison between symmetric and non-symmetric wave functions, we define ${M^2_{\text{str}}=\tfrac{1}{4N}\sum_{\alpha}S(\boldsymbol{k}_{\alpha})}$ where $\boldsymbol{k}_{\alpha}$ denote the four incommensurate stripe-ordering wavevectors. Their explicit values, together with additional details, are provided in \cref{sec:spin_structure_factor} of the \textit{Appendix}. All correlation functions are evaluated assuming translational invariance by averaging over lattice translations. In both panels, for each architecture, we report three points corresponding to optimization without symmetry projection (circles), with translational symmetry restored (rhombi), and with both translational and rotational symmetries restored (triangles).

These results clearly demonstrate the impact of architectural biases on the resulting physical properties: following the initial optimization and before symmetry restoration, determinant ans\"atze in the standard and particle–hole representations converge to qualitatively different solutions. The backflow determinant exhibits a relatively weak $d$-wave order parameter, approximately four times smaller than in the particle–hole representation, while displaying stronger stripe magnetic order. By contrast, the particle–hole construction enhances superconducting correlations and suppresses spin correlations. The Pfaffian backflow state yields intermediate values for both order parameters, interpolating between these two limits. Refer to \cref{sec:antisymm_schemes} of the \textit{Appendix} for further details about the biases of the three antisymmetrization schemes at mean field level. 

The qualitatively different solutions obtained before restoring the symmetries lie extremely close in the energy landscape. 
\Cref{table:energies} summarizes the ground-state variational energies for the three antisymmetrization schemes. For each architecture, we report the energy and variance per site obtained without symmetry projection and after restoring translational and rotational symmetries. A zero-variance extrapolation provides an estimate of the exact ground-state energy per site, $E_0=-0.7540(1)$ (see \cref{sec:zero_variance} of the \textit{Appendix} for more details). Using this reference, we define the relative energy error $\Delta\varepsilon={|E_{\theta}-E_0|}/{|E_0|}$ (with $E_{\theta}$ the variational energy per site), which is reported in the last column of \cref{table:energies}. The first row lists the best previous result from Ref.~\cite{gu2025}, obtained with a variational wave function closely related to our Transformer with determinant backflow, with differences regarding the implementation of the Attention Mechanism~\cite{viteritti2023prl, rende2024prr, viteritti2026approaching}, refer to \cref{sec:transformer} of the \textit{Appendix} for more details. The energy reported in Ref.~\cite{gu2025} corresponds to an estimated accuracy of $10^{-2}$, comparable to our determinant-based results before symmetry restoration. In contrast to Ref.~\cite{gu2025}, however, all optimizations here are performed from random initialization, without mean-field pretraining or auxiliary stabilization procedures. Importantly, the three solutions obtained in this work, even before symmetry restoration, are lower than the best precedent result; yet, they correspond to qualitatively different physical states. 
Upon systematic variance reduction toward an unbiased solution via symmetry projection, all antisymmetrization schemes converge toward compatible values of spin and pairing order (see \cref{fig:order_par} and \cref{table:energies}).
The Pfaffian-based backflow state achieves the best variational energy value $-0.75058(1)$ corresponding to a relative error of $4.5\times10^{-3}$, nearly a factor-of-three improvement over the best previous result from Ref.~\cite{gu2025}.

%%%%%%%%%%%%%%%%%%%%%%%%%%%%%%%%
\begin{table}[t]
\centering
\begin{tabular}{|c|c|c| c|c|}
\hline\hline
\textbf{Ansatz} & \textbf{Symm} & \textbf{Energy} & \textbf{Variance} & \textbf{Rel. Error} \\
\hline
\makecell{Ref.~\cite{gu2025}} & - & $-0.7447$ & -- & $0.012$ \\
\hline
\multirow{3}{*}{Det} 
& - & -0.74512(1) & 0.037 & $0.012$ \\
& $T$ & -0.74834(1) & 0.024 & $0.0081$ \\
& $T{+}R$ & -0.74993(1) & 0.019 & $0.0056$ \\
\hline
\multirow{3}{*}{Det-PH} 
& - & -0.74649(1) & 0.037 & $0.010$ \\
& $T$ & -0.74934(1) & 0.022 & $0.0062$ \\
& $T{+}R$ & -0.75026(1) & 0.017 & $0.005$ \\
\hline
\multirow{3}{*}{Pfaffian} 
& - & -0.74705(1) & 0.037 & $0.0092$ \\
& $T$ & -0.74952(1) & 0.021 & $0.0059$ \\
& $T{+}R$ & \textbf{-0.75058(1)} & \textbf{0.016} & \textbf{0.0045}  \\
\hline
\makecell{Zero-Var.Extr.} 
& - & -0.7540(4) & - & - \\
\hline\hline
\end{tabular}
\caption{\label{table:energies}
Ground-state energy and variance per site for different variational states of the Hubbard model [see \cref{eq:Hubbard}] on an $8\times 8$ cluster with PBC, at $U=8.0$, $t^{\prime}=-0.2$, and doping $\delta=1/8$. For each {\it ansatz}, we report results obtained without symmetry projection ($-$), with translational symmetry  ($T$), and with both translational and rotational symmetries ($T+R$) restored. The relative error (last column) is defined as $\Delta\varepsilon = {|E_{\theta} - E_0|}/{|E_0|}$, with $E_{\theta}$ the variational energy per site and $E_0=-0.7540(1)$ is obtained from zero-variance extrapolation (refer to \cref{sec:zero_variance} of the \textit{Appendix} for additional details). The fully symmetrized Pfaffian is highlighted in bold, as it yields the best variational result.}
\end{table}
%%%%%%%%%%%%%%%%%%%%%%%%%%%%%%%%

%%%%%%%%%%%%%%%%%%%%%%%%%%%%%%%%
\begin{figure}[t]
    \begin{center}
\centerline{\includegraphics[width=\columnwidth]{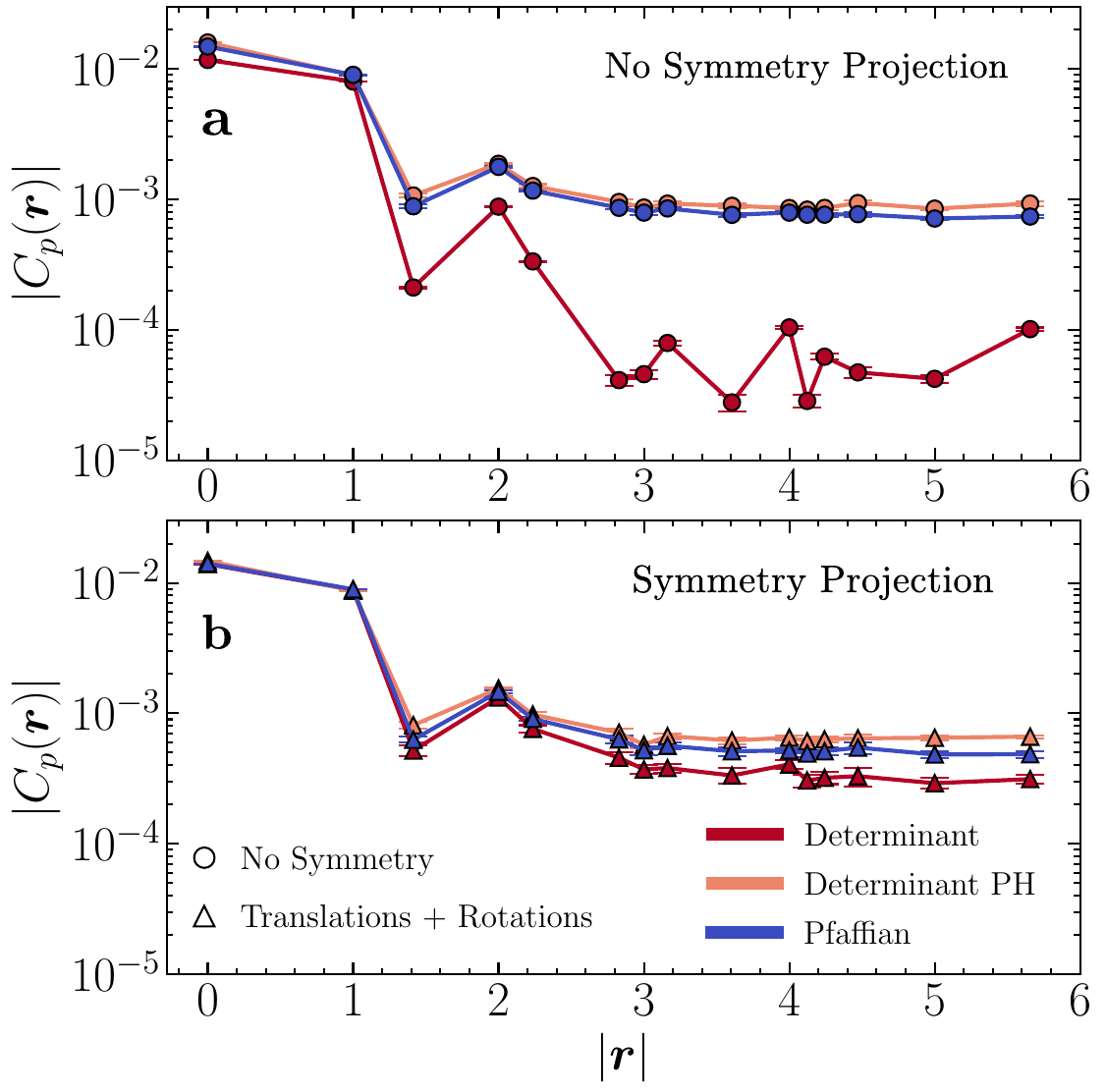}}
        \caption{\label{fig:pairing} Absolute value of the $d$-wave pairing correlations $|C_p(\boldsymbol{r})|$ [see \cref{eq:pairing_corr}] as a function of distance $|\boldsymbol{r}|$ on an $8\times8$ square lattice with PBC at $U=8.0$, $t^{\prime}=-0.2$, and $\delta=1/8$. \textbf{Panel a.} Results without symmetry projection. \textbf{Panel b.} Results after restoring translational and rotational symmetries.}
    \end{center}
\end{figure}
%%%%%%%%%%%%%%%%%%%%%%%%%%%%%%%%

We stress that, once sufficient variational accuracy is reached, the different variational states converge to the same underlying physics. Importantly, if the procedure is halted at the initial stage, i.e., before symmetry restoration and systematic variance reduction, the resulting state can remain strongly biased and fail to faithfully represent the correct physical picture. In this regime, architectural biases may stabilize quasi-degenerate solutions dominated by a specific order. A concrete example is provided by Ref.~\cite{gu2025}, where a backflow determinant-based {\it ansatz} without symmetry restoration yields states with robust stripe order across a broad range of system sizes and parameter regimes of the Hubbard model. Our results indicate that such outcomes may reflect the intrinsic bias of the chosen antisymmetrization rather than a definitive characterization of the ground state.

%%%%%%%%%%%%%%%%%%%%%%%%%%%%%%%%
\begin{figure*}[ht]
    \begin{center}
\centerline{\includegraphics[width=2.1\columnwidth]{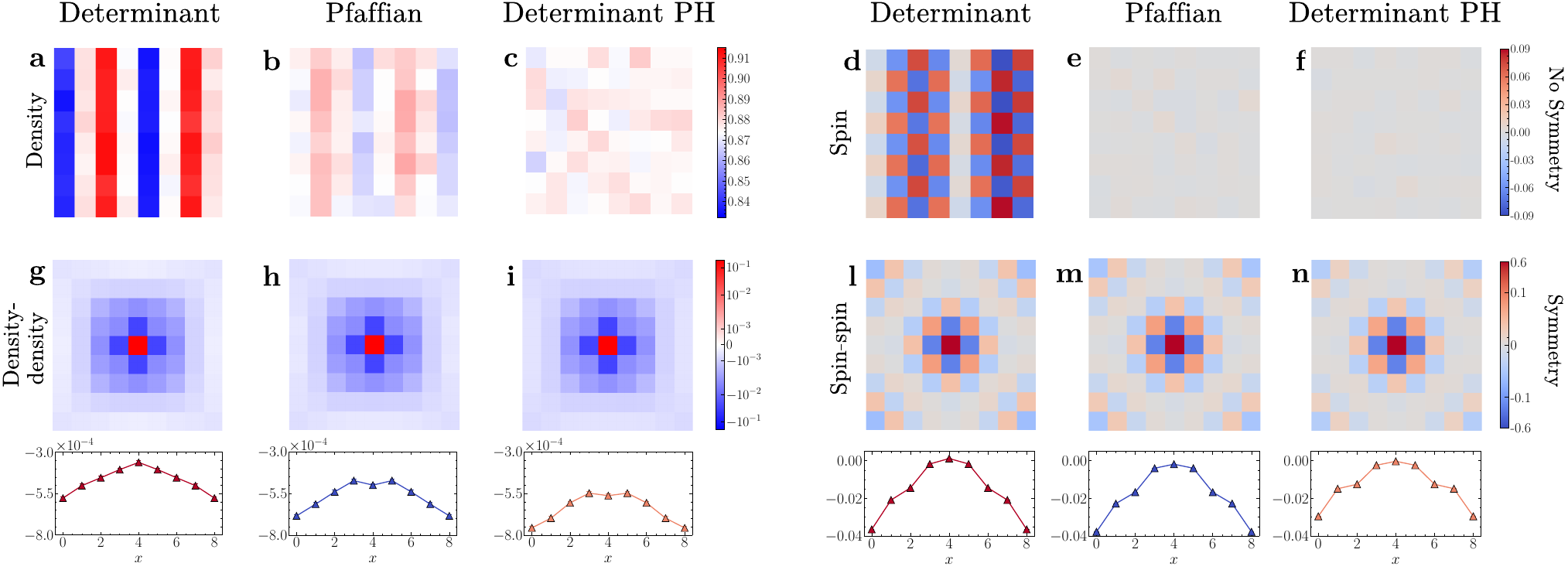}}
        \caption{\label{fig:stripe_order} Real-space charge density and spin patterns for the $8\times 8$ Hubbard model with PBC at $U=8.0$, $t^{\prime}=-0.2$, and ${\delta=1/8}$ [see \cref{eq:Hubbard}]. Columns compare the three variational constructions: determinant, Pfaffian, and determinant in the PH representation. The top row shows results without symmetry projection, while the bottom row shows the corresponding symmetry-projected results. \textbf{Panels a,b,c}: Charge density $N(\boldsymbol r)$ without symmetry projection. \textbf{Panels d,e,f}: Local spin $\langle \hat{S}^z_{\boldsymbol r}\rangle$ without symmetry projection. \textbf{Panels g,h,i}: Density-density correlations $C_D(\boldsymbol r)$ after symmetry projection. \textbf{Panels l,m,n}: Spin-spin correlations $C_S(\boldsymbol r)$ after symmetry projection.
        Below each panel we show the one-dimensional cuts of the density $C_D(x,y=4)$ and the staggered spin correlation $(-1)^x C_S(x,y=4)$. Definitions of all observables are given in the main text.}
    \end{center}
\end{figure*}
%%%%%%%%%%%%%%%%%%%%%%%%%%%%%%%%

\subsection{Pairing order}\label{sec:pairing}
\Cref{fig:pairing} reports the absolute value of the pairing correlation functions $|C_p(\boldsymbol{r})|$ [see \cref{eq:pairing_corr}] as a function of the distance $|\boldsymbol{r}|$ in the lattice for the three antisymmetrization schemes. In panel (a), we report the results before symmetry projections. The determinant backflow {\it ansatz} exhibits a pronounced suppression of long-range pairing correlations. For $|\boldsymbol{r}|\gtrsim 3$, the signal is nearly one order of magnitude smaller with the respect to the other states. The determinant-PH backflow state alleviates this suppression, providing a solution closer to the Pfaffian {\it ansatz}. Enforcing the symmetries, alongside the quantitative improvement in energy and variance (see \cref{table:energies}), we observe a substantially greater consistency in the pairing profiles across architectures, as shown in panel (b) of \cref{fig:pairing}. In particular, the Pfaffian and determinant-PH pairing amplitudes slightly decrease, whereas the pure determinant result increases by nearly an order of magnitude, remaining only moderately smaller relative to the other two solutions (refer to panel (a) of \cref{fig:order_par} for the corresponding order parameters). 

For additional analysis about the pairing correlation functions refer to \cref{sec:pairing_sign} of the \textit{Appendix}.

\subsection{Stripe order} 
We next analyze stripe order, which recent studies suggest may coexist with superconductivity~\cite{xu2024, Roth2025SCNQS}. In \cref{fig:stripe_order}, we compare the real-space charge density and spin patterns. The upper panels report the one-body observables obtained without symmetry projection: the local charge density,
$N(\boldsymbol{r})=\langle n_{\boldsymbol{r}\uparrow}+n_{\boldsymbol{r}\downarrow}\rangle$,
shown in panels (a)–(c), and the local spin,
$\langle \hat{S}^z_{\boldsymbol{r}} \rangle$,
shown in panels (d)–(f), for the three variational ans\"atze. Among the unprojected states, the determinant-backflow {\it ansatz} displays strong spatial modulations in both charge and spin, indicating a pronounced tendency toward stripe formation. By contrast, the determinant-PH and Pfaffian states yield significantly more uniform profiles. These differences indicate that the initial optimization can converge to distinct solutions, even though their variational energies remain very close (see \cref{table:energies}).

The lower panels show the corresponding two-body correlation functions after full symmetry projection: the density-density correlation $C_D(\boldsymbol{r})=\langle \hat{n}_{\boldsymbol{0}}\hat{n}_{\boldsymbol{r}}\rangle - n^2$,
in panels (g)–(i), and the spin-spin correlation $C_S(\boldsymbol{r})=\langle \hat{S}^z_{\boldsymbol{0}}\hat{S}^z_{\boldsymbol{r}}\rangle$,
in panels (l)–(n). After restoring translational and rotational symmetries, the results obtained from the different antisymmetrization schemes become remarkably consistent. In particular, both the charge and spin correlations exhibit the same spatial structure, showing that the discrepancies seen before projection are largely removed once the variational state is refined. This convergence is further confirmed by the one-dimensional cuts of $C_D(x,y=4)$ and of the staggered spin correlation $(-1)^x C_S(x,y=4)$ shown below each panel.

Stripe order is most clearly visible in the spin-spin correlations. In addition to the antiferromagnetic pattern, $C_S(\boldsymbol{r})$ exhibits a longer-wavelength modulation that weakens the antiferromagnetic order, consistent with the additional peaks observed in the spin structure factor (see \cref{sec:spin_structure_factor} of the \textit{Appendix} for further details). 
By contrast, for the present system size we only find very weak charge modulation in the density-density correlations. We expect that charge modulations may become more visible only on larger clusters~\cite{Roth2025SCNQS}, a detailed investigation is beyond the scope of the present work and is left for future study.

Overall, these results show that the marked differences observed before symmetry restoration, especially the pronounced stripe pattern found in the determinant-based {\it ansatz}, do not correspond to distinct physical phases. Rather, they reflect an {\it ansatz}-dependent bias that persists at finite variational accuracy.

\section{Attractive Hubbard model}
Finally, we further support the main conclusion of this work by considering the attractive Hubbard model, where $s$-wave superconductivity is established by numerically exact quantum Monte Carlo calculations. We focus on the case $t'=0.0$ and $U=-8.0$, at hole doping $\delta=1/8$, on an $8\times 8$ square lattice with periodic boundary conditions. In this regime, a numerically exact benchmark is provided by Determinant Quantum Monte Carlo (DQMC)~\cite{blankenbecler1981, hirsch1985, white1989, scalapino1993}.

%%%%%%%%%%%%%%%%%%%%%%%%%%%%%%%%
\begin{figure}[t]
    \begin{center}
\centerline{\includegraphics[width=\columnwidth]{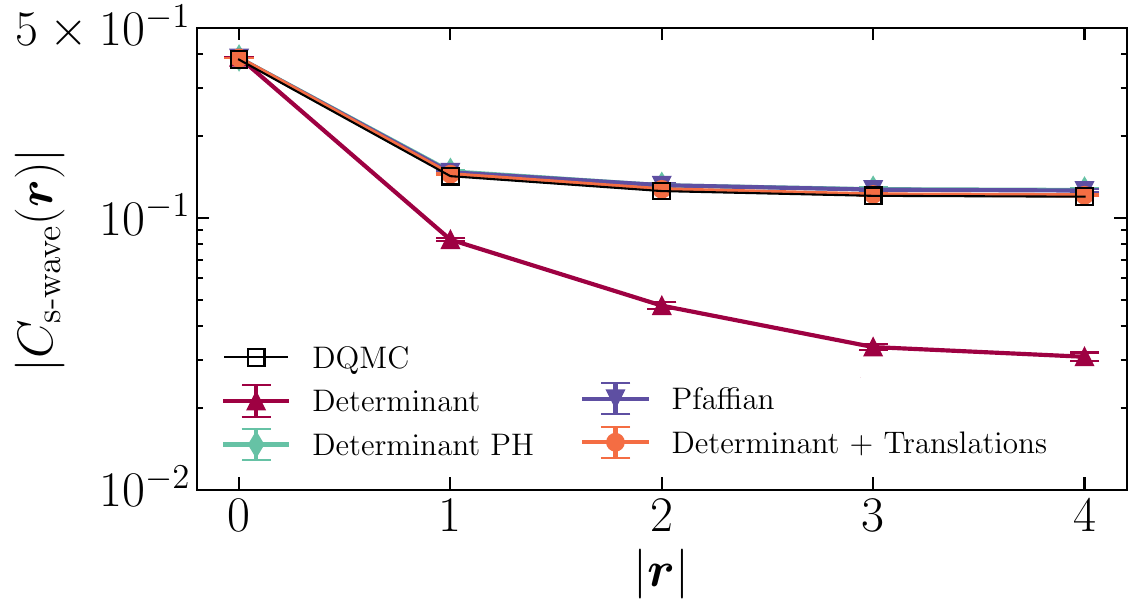}}
        \caption{\label{fig:swave} $s$-wave pairing correlation function $C_{\text{s-wave}}(\boldsymbol{r})$ [see \cref{eq:s_wave}] for the attractive Hubbard model with $U=-8.0$, ${t'=0.0}$, and $\delta=1/8$ on an $8\times8$ lattice along the diagonal of the square. Results are shown for determinant, determinant in PH and Pfaffian, symmetry-restored determinant {\it ansatz}, and DQMC.}
    \end{center}
\end{figure}
%%%%%%%%%%%%%%%%%%%%%%%%%%%%%%%%

To characterize superconducting order, we compute the $s$-wave pair-pair correlation function
\begin{equation}\label{eq:s_wave}
    {C_{s\text{-wave}}(\boldsymbol{r})=
\langle
\hat{c}^{\dagger}_{\boldsymbol{r}\downarrow}
\hat{c}^{\dagger}_{\boldsymbol{r}\uparrow}
\hat{c}_{\boldsymbol{0}\uparrow}
\hat{c}_{\boldsymbol{0}\downarrow}
\rangle} \ ,
\end{equation}
whose large-distance behavior signals the presence of superconducting order. 

In \cref{fig:swave} we report $C_{s\text{-wave}}(\boldsymbol{r})$ as a function of $|\boldsymbol{r}|$ along the diagonal of the $L\times L$ cluster. Even in this comparatively simple setting, the determinant-based construction substantially underestimates the pairing correlations with respect to the DQMC benchmark by one order of magnitude. This result reinforces the general picture emerging from the present study: enforcing antisymmetry through a determinant output layer, even when supplemented by backflow, introduces a significant bias in the description of superconductivity and tends to suppress the pairing signal. Only after restoring translational symmetry do the corresponding results come into very close agreement with the exact benchmark. By contrast, the Pfaffian and particle-hole determinant ans\"atze already provide a much more robust and physically consistent description of pairing correlations without symmetry restoration. 

This qualitative difference is also reflected quantitatively in \cref{table:energies_singlecol}, where we report the energies of the different variational states. In particular, translational symmetry restoration improves the accuracy of the determinant by about one order of magnitude. Although the determinant still does not reach the same accuracy as the Pfaffian and particle-hole determinant states, beyond a sufficient level of variational accuracy the correlation functions obtained from the different ans\"atze become physically compatible. This mirrors the behaviour found in the repulsive case and supports the conclusion that agreement in the physical properties emerges once the wave function is sufficiently accurate.

\section{Conclusions}
We have examined how different backflow implementations affect the variational description of physical properties in the two-dimensional Hubbard model. Using a common Transformer-based backflow architecture and optimizing all states from random initial conditions, we find that different antisymmetrization schemes can initially converge to qualitatively distinct states, ranging from stripe-dominated to strongly superconducting, despite nearly degenerate energies.

On the one hand, our results show that the choice of antisymmetrization scheme can bias the correlation functions obtained from neural-network fermionic wave functions towards phases that are naturally encoded at the mean-field level. On the other hand, we demonstrate that this bias can be overcome by increasing variational accuracy through symmetry restoration, after which all the ans\"atze yield similar correlation functions. Our findings highlight the importance of tracking how correlation functions evolve as the wave function is systematically refined under variance reduction.

%%%%%%%%%%%%%%%%%%%%%%%%%%%%%%%%
\begin{table}[t]
\centering
\begin{tabular}{|c| c| c |c |c|}
\hline\hline
\textbf{Ansatz} & \textbf{Symm} & \textbf{Energy} & \textbf{Variance} & \textbf{Rel. Error} \\
\hline
\multirow{2}{*}{Det} 
& - & -3.9950(1) & 0.051 & $0.0055$ \\
& $T$ & -4.0152(1) & 0.017 & 0.00044 \\
\hline
\makecell{Det-PH} & - & -4.0167(1) & $0.0024$ & $0.000075$ \\
\hline
\makecell{Pfaffian} & - & -4.0169(1) & $0.0014$ & $0.000024$ \\
\hline
\makecell{DQMC} & - & -4.017 & - & - \\
\hline\hline
\end{tabular}
\caption{\label{table:energies_singlecol}
Benchmark table for the pure Hubbard model ($t'=0$) with attractive interaction $U=-8.0$ at hole doping $\delta=1/8$ on an $8\times 8$ lattice. We report results obtained without symmetry projection ($-$) and with translational symmetry projection ($T$) for the determinant. The relative error (last column) is defined as $\Delta\varepsilon = {|E_{\theta} - E_{\text{DQMC}}|}/{|E_{\text{DQMC}}|}$, with $E_{\theta}$ the variational energy per site and $E_{\text{DQMC}}=-4.017$ is the DQMC energy.}
\end{table}
%%%%%%%%%%%%%%%%%%%%%%%%%%%%%%%%
These observations are particularly relevant in the broader context of the square-lattice Hubbard model, which is widely regarded as a minimal model for capturing key aspects of cuprate phenomenology, in particular superconductivity. Our results, therefore, provide a cautionary message: variational architectures that cannot naturally accommodate off-diagonal long-range order, such as the backflow determinant without symmetry restoration considered here, may achieve very competitive energies while still missing superconducting correlations. In such cases, great care must be taken to account for this bias.

In light of our findings, we propose the $8\times 8$ Hubbard model on a square lattice with periodic boundary conditions, at $t'=-0.2$, $U=8$, and $1/8$ hole doping, as a challenging setup for benchmarking numerical methods targeting correlated fermionic systems. The close energetic competition among distinct orders, combined with the strong sensitivity of the correlation functions to the choice of \textit{ansatz}, makes this parameter set a stringent test of any numerical approach. Reproducing the physical picture established here with independent techniques, such as tensor network methods~\cite{liu2025}, auxiliary-field quantum Monte Carlo~\cite{hao2022,ryan2024}, or other neural-network-based approaches, is an important step toward consolidating our understanding of this regime. More broadly, establishing the ground-state properties accurately and unambiguously on finite clusters of this size is a necessary prerequisite before reliable conclusions can be drawn about the physics of the Hubbard model on larger systems and in the thermodynamic limit.

At the same time, our results also provide a hopeful perspective. They suggest that sufficiently accurate and flexible variational approaches, potentially leveraging the large expressivity of deep neural networks, can overcome these {\it ansatz}-dependent biases and recover a consistent physical picture. This opens a promising route towards a controlled understanding of the competing phases of the Hubbard model and their connection to high-$T_c$ cuprate superconductivity.

\section*{Appendix}
\appendix

\subsection{Transformer Backflow Architecture}\label{sec:transformer}
The Transformer Backflow Architecture takes as input a physical configuration $\boldsymbol{s}=({s}_1, \dots, {s}_N)$, where $s_i \in \{0,1,2,3\}$ corresponds to the possible four states of the local Hilbert space of site $i$: empty, singly occupied spin-$\uparrow$ or spin-$\downarrow$, and doubly occupied site. Following the standard tokenization scheme~\cite{gu2025, rende2026alm}, each input $s_i$ is mapped to an embedding vector $\boldsymbol{x}_i \in \mathbb{R}^d$, yielding the input sequence of the neural network $(\boldsymbol{x}_1, \ldots, \boldsymbol{x}_N)$, where $d$ is the embedding dimension. The network outputs a new sequence of vectors $(\boldsymbol{y}_1, \ldots, \boldsymbol{y}_N)$ with $\boldsymbol{y}_i \in \mathbb{R}^d$~\cite{rende2026alm}. The Transformer architecture used in this work has $n_l = 4$ layers, $h = 12$ attention heads, and an embedding dimension of $d = 72$ (refer to Ref.~\cite{viteritti2025prb} for more details about their role). 
%%%%%%%%%%%%%%%%%%%%%%%%%%%%%%%%
\begin{figure}[t]
    \begin{center}
\centerline{\includegraphics[width=0.8\columnwidth]{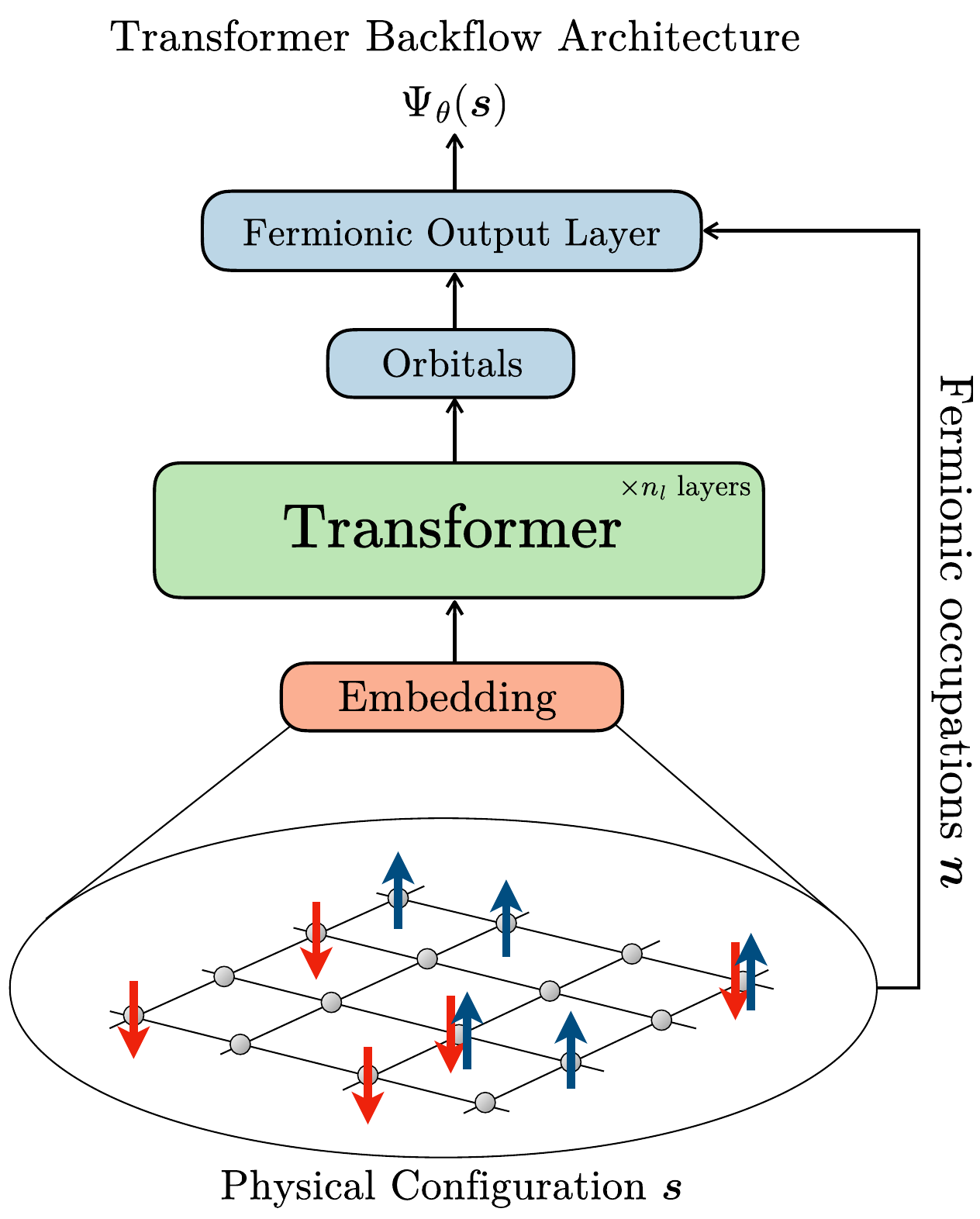}}
\caption{\label{fig:architecture} In the Transformer Backflow Architecture a physical configuration $\boldsymbol{s}$ (see definition in the main text) is mapped through an Embedding layer into a sequence of vectors, which are then processed by a stack of $n_l$ Transformer layers. The Transformer output is fed into a Fermionic Output Layer that constructs configuration-dependent orbitals. These orbitals, together with the fermionic occupation numbers $\boldsymbol{n}$, are combined via the chosen antisymmetrization scheme (determinant or Pfaffian) to produce the wavefunction amplitude $\Psi_\theta(\boldsymbol{s})$.}
    \end{center}
\end{figure}
%%%%%%%%%%%%%%%%%%%%%%%%%%%%%%%%
Relative to previous Transformer-backflow implementations~\cite{sharma2025,gu2025,ma2025}, the architecture incorporates a Factored Attention Mechanism~\cite{rende2024prr,viteritti2023prl} with a spatial inductive bias~\cite{viteritti2026approaching}. These components, introduced for two-dimensional frustrated spin systems, enable stable and accurate simulations on large lattices~\cite{viteritti2026approaching}.

For determinant-based antisymmetrization, the Transformer output defines effective single-particle orbitals via a site-resolved linear map~\cite{diluo_prl, gu2025, rende2026alm}:
\begin{equation}\label{eq:backflow_orbitals}
\Phi_{i\sigma\alpha} = \sum_{\beta=1}^d y_{i\beta} W_{i\sigma\alpha\beta},
\end{equation}
where $W_{i\sigma\alpha\beta}$ are trainable parameters and $\alpha = 1, \ldots, N_e$ labels the orbitals, with $N_e$ being the number of fermions. Introducing the composite index ${r = (i,\sigma)}$, the tensor is reshaped into $\Phi_{r\alpha} \in \mathbb{R}^{2N \times N_e}$. The wavefunction amplitude is obtained as 
\begin{equation}\label{eq:det_ansatz}
\Psi_{\theta}(\boldsymbol{s}) = \det \left[ \boldsymbol{n} \star \Phi(\boldsymbol{s}) \right] \ ,
\end{equation}
where $\boldsymbol{n} \star \Phi(\boldsymbol{s}) $ denotes the $N_e \times N_e$ submatrix of $\Phi(\boldsymbol{s}) $ indexed by the occupation numbers of ${\boldsymbol{n}=(n_{1\uparrow},\dots, n_{N\uparrow},n_{1\downarrow},\dots, n_{N\downarrow})}$. 

For the Pfaffian scheme, the same construction used in \cref{eq:backflow_orbitals} is used to generate a matrix ${\phi_{r\alpha} \in \mathbb{R}^{2N \times 2N}}$. These orbitals are then paired through an antisymmetric (configuration independent) matrix ${A \in \mathbb{R}^{2N \times 2N}}$ of trainable parameters~\cite{gao2024, loehr2025, Roth2025SCNQS}. The resulting wavefunction amplitude is
\begin{equation}
\label{eq:pfaffian_ansatz}
\Psi_{\theta}(\boldsymbol{s}) = \mathrm{Pf} \left[ \boldsymbol{n} \star \phi(\boldsymbol{s})\  A \ \phi(\boldsymbol{s})^T \star \boldsymbol{n} \right] \ .
\end{equation}
Refer to \cref{fig:architecture} for a schematic representation of the Transformer backflow architecture.

The spatial symmetries of Hubbard Hamiltonian [see \cref{eq:Hubbard}] can be enforced on the states in \cref{eq:det_ansatz} and \cref{eq:pfaffian_ansatz} through quantum-number projection~\cite{mizusaki2004, tahara2008, nomura2021iop, viteritti2022, reh2023}. Given a symmetry group $G$ with $|G|$ elements, we define the projected variational state
\begin{equation}
\Psi_{\lambda}(\boldsymbol{s})= \frac{1}{|G|}
\sum_{g\in G}
\chi_{\lambda}^{*}(g)\,
\xi_{g^{-1}}(\boldsymbol{s})\,
\Psi_{\theta}(g^{-1}\boldsymbol{s}) \ ,
\end{equation}
where $\lambda$ labels the irreducible representation and $\chi_{\lambda}(g)$ is its character. In contrast to spin systems~\cite{nomura2021iop,rende2024stochastic,chen2024empowering}, in fermionic systems one must also include the permutation sign $\xi_{g^{-1}}(\boldsymbol{s})$, which depends on the fixed ordering of the sites~\cite{robledo2022,sharma2025,loehr2025,rende2026alm}. Further details can be found in Ref.~\cite{sharma2025}. 

For the repulsive Hubbard model considered here, we implement lattice translations with momentum $\boldsymbol{k}=(0,0)$ and the $C_4$ point-group symmetry with eigenvalue $-1$.

\subsection{Mean Field Biases of Fermionic Ans\"atze}\label{sec:antisymm_schemes}
In this Section, we discuss the mean-field biases associated with the three fermionic ans\"atze used in this work: the determinant, the determinant in the particle-hole representation, and the Pfaffian. Throughout, we consider the mean-field limit, in which the orbitals are taken to be configuration independent.

The Slater-determinant wavefunction is defined as
\begin{equation}
    \Psi_{\text{det}}(\boldsymbol{n}) = \det\left[U\star {\boldsymbol{n}}\right]
\end{equation}
where $U\in\mathbb{R}^{2N\times N_e}$ is a matrix of variational parameters. In the non-interacting limit, the determinant {\it ansatz} corresponds to the ground state of the quadratic Hamiltonian
\begin{equation}\label{eq:mf_ham}
\hat{H}_{\text{det}}
=
-\sum_{i,j,\sigma} t_{ij}\, \hat{c}_{i\sigma}^{\dagger} \hat{c}_{j\sigma}
-\mu \sum_{i,\sigma} \hat{n}_{i\sigma}
+\sum_{i,j,\sigma,\sigma'} m_{ij}^{\sigma\sigma'}\hat{c}_{i\sigma}^{\dagger}\hat{c}_{j\sigma'} \ ,
\end{equation}
where $t_{ij}$ denotes the hopping amplitudes, $\mu$ the chemical potential, and $m_{ij}^{\sigma\sigma'}$ the spin-dependent terms, including both longitudinal and transverse magnetic contributions. This Hamiltonian can therefore accommodate magnetic fluctuations, but cannot describe superconducting pairing.

The simplest superconducting terms that can be added at the mean-field level are singlet pairing terms. To this end, consider the Hamiltonian
\begin{equation}\label{eq:ph_ham}
\begin{split}
\hat{H}_{\text{PH}}
&=
-\sum_{i,j,\sigma} t_{ij}\,\hat{c}_{i\sigma}^{\dagger}\hat{c}_{j\sigma}
-\mu \sum_{i,\sigma} \hat{n}_{i\sigma} + \sum_{i,j,\sigma} m_{ij}^{\sigma\sigma}\hat{c}_{i\sigma}^{\dagger}\hat{c}_{j\sigma} \\
&\quad
+\sum_{i,j}
\left[
\Delta_{ij}
\left(
\hat{c}_{i\uparrow}^{\dagger}\hat{c}_{j\downarrow}^{\dagger}
-
\hat{c}_{i\downarrow}^{\dagger}\hat{c}_{j\uparrow}^{\dagger}
\right)
+\mathrm{h.c.}
\right] \ .
\end{split}
\end{equation}
Under a particle-hole transformation $c_{i \uparrow} \rightarrow c_{i \uparrow}$, ${c_{i \downarrow} \rightarrow c^\dagger_{i \downarrow}}$ the singlet pairing terms are mapped to spin-mixing, number-conserving terms. As a result, $\hat{H}_{\mathrm{PH}}$ [see \cref{eq:ph_ham}] can be rewritten as a quadratic Hamiltonian of the same general form as $\hat{H}_{\mathrm{det}}$ [see \cref{eq:mf_ham}]. Its ground state can therefore be represented as a determinant in the particle-hole basis. The advantage of this construction is that it allows superconducting correlations to be represented within a determinant {\it ansatz}.

However, to include magnetic terms, both longitudinal and transverse, together with superconducting terms in both the singlet and triplet channels, one must instead consider the most general quadratic Hamiltonian:
\begin{equation}
\label{eq:pf_ham}
\begin{split}
\hat{H}_{\mathrm{Pf}}
&=
-\sum_{i,j,\sigma} t_{ij}\,\hat{c}_{i\sigma}^{\dagger}\hat{c}_{j\sigma}
-\mu \sum_{i,\sigma} \hat{n}_{i\sigma} \\
&\quad
+\sum_{i,j,\sigma,\eta} m_{ij}^{\sigma\sigma'}\,\hat{c}_{i\sigma}^{\dagger}\hat{c}_{j\sigma'}
+\frac{1}{2}\sum_{i,j,\sigma,\eta}
\left(
\Delta_{ij}^{\sigma\sigma'}\,\hat{c}_{i\sigma}^{\dagger}\hat{c}_{j\sigma'}^{\dagger}
+\mathrm{h.c.}
\right) \, .
\end{split}
\end{equation}
In general, the ground state of this Hamiltonian cannot be written as a single Slater determinant and instead requires a Pfaffian representation of the form
\begin{equation}
    \Psi_{\text{Pf}}(\boldsymbol{n})
    =
    \text{Pf}\!\left[\boldsymbol{n}\star V A V^{T}\star \boldsymbol{n}\right] \, ,
\end{equation}
where $V\in\mathbb{R}^{2N\times 2N}$ is a general matrix and $A\in\mathbb{R}^{2N\times 2N}$ is antisymmetric. Therefore, at the mean-field level, the Pfaffian {\it ansatz} can simultaneously capture magnetic and superconducting fluctuations.

As discussed in \cref{sec:results} of the main text, in the absence of translational symmetry restoration, the backflow Slater determinant converges to a state with predominantly magnetic order, whereas the particle-hole determinant converges to a state with predominantly superconducting correlations. By contrast, the backflow Pfaffian yields an intermediate state in which both types of order are present already before symmetry projection. The mean-field analysis presented here reproduces the same qualitative tendencies observed in the full backflow ans\"atze, showing that the choice of antisymmetrization scheme induces a strong bias even when supplemented by a highly expressive backflow transformation. As shown in \cref{sec:results}, these biases can nevertheless be reduced by systematically improving the variational accuracy, for example, through symmetry restoration.

%%%%%%%%%%%%%%%%%%%%%%%%%%%%%%%%
\begin{figure}[t]
    \begin{center}
\centerline{\includegraphics[width=\columnwidth]{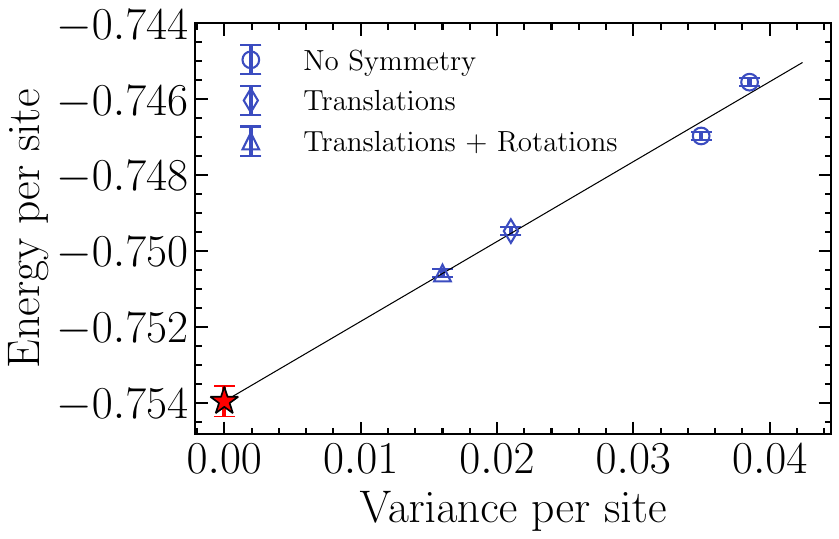}}
        \caption{\label{fig:zero_variance} Variational energy per site as a function of the energy variance per site for the $8\times 8$ Hubbard model with periodic boundary conditions at $U=8.0$, $t'=-0.2$, and doping $\delta=1/8$. Results are shown for the Pfaffian state with different levels of accuracy obtained without symmetry projection (circles), with translational symmetry (rhombi), and with both translational and rotational symmetries (triangles) restored. A linear fit of the data is used to extrapolate to zero variance and estimate the ground-state energy (red star).}
    \end{center}
\end{figure}
%%%%%%%%%%%%%%%%%%%%%%%%%%%%%%%%

\subsection{Zero-Variance Extrapolation}\label{sec:zero_variance}
An estimate of the exact ground-state energy for the $8\times 8$ Hubbard model at $U=8.0$, $t'=-0.2$, and doping $\delta=1/8$ can be obtained by means of zero-variance extrapolation~\cite{becca2013, becca2015, viteritti2026approaching}. We perform this analysis using the Pfaffian {\it ansatz}, since, among the variational states considered, it provides the most accurate description, as also indicated by the variational energies reported in \cref{table:energies}. In \cref{fig:zero_variance}, we report the variational energy per site
$E_{\theta}={\langle \Psi_{\theta}| \hat{H}|\Psi_{\theta} \rangle}/{N}$, as a function of the energy variance per site
$\sigma^2_{\theta}=[{\langle \Psi_{\theta}| \hat{H}^2|\Psi_{\theta} \rangle-\langle \Psi_{\theta}| \hat{H}|\Psi_{\theta} \rangle^2}]/N$, at different stages of the optimization and symmetry restoration. In particular, we consider states obtained after $10^4$ and $4\times 10^4$ optimization steps, as well as after restoring translational and rotational symmetries. For sufficiently accurate variational states, the energy is expected to depend linearly on the variance according to
$E_{\theta} \approx E_0 + \mathrm{const}\times \sigma^2_{\theta}$,
where $E_0$ denotes the exact ground-state energy.

As shown in \cref{fig:zero_variance}, the data exhibit an approximately linear behavior over the range of variances considered. By performing a linear fit of the energy as a function of the variance, we extract the zero-variance intercept and obtain $E_0=-0.7540(1)$, as reported in \cref{table:energies}. The uncertainty accounts for both the statistical errors of the data points and the stability of the extrapolation upon varying the fitting range.

%%%%%%%%%%%%%%%%%%%%%%%%%%%%%%%%
\begin{figure}[t]
    \begin{center}
\centerline{\includegraphics[width=\columnwidth]{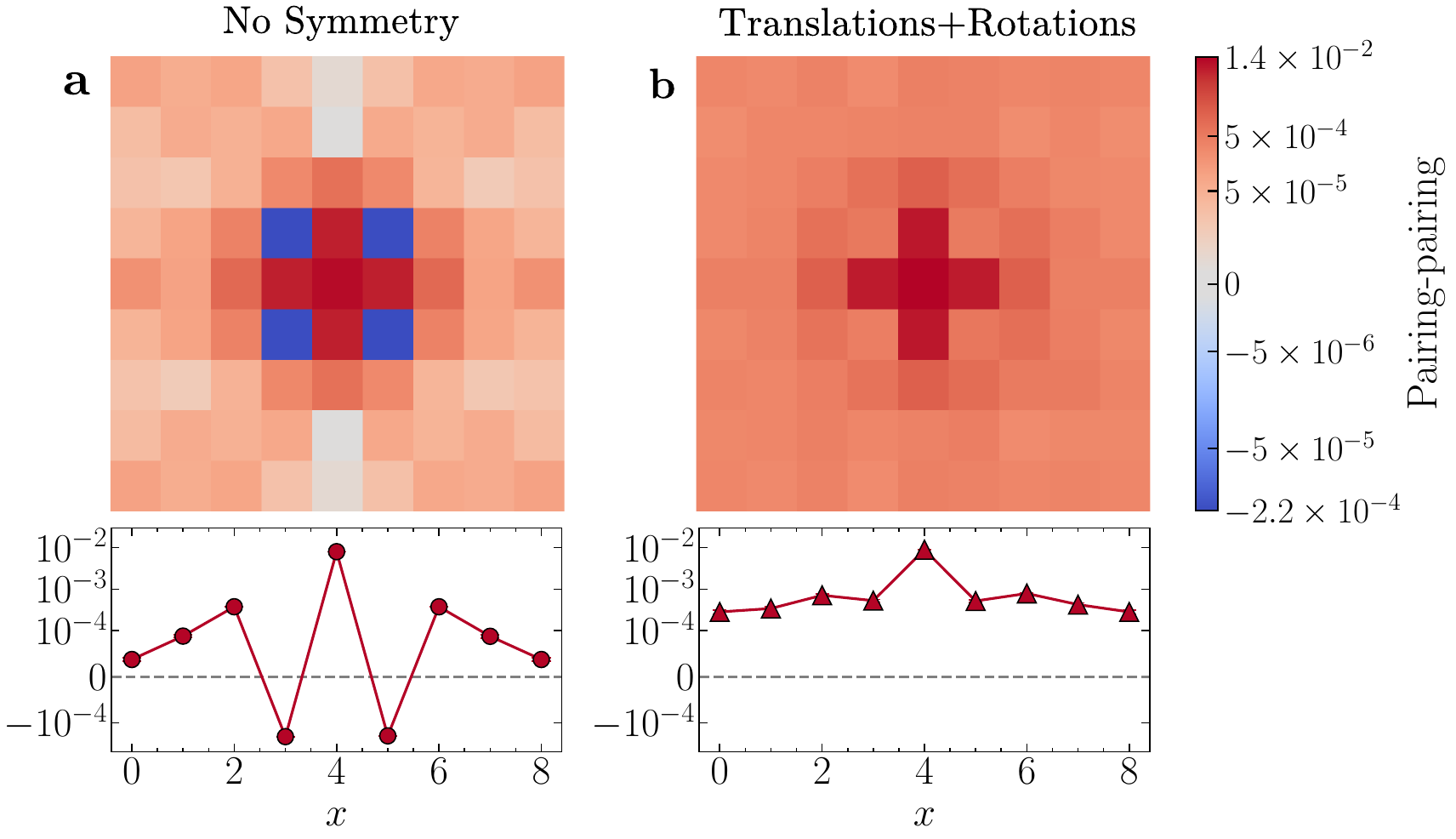}}
        \caption{\label{fig:pairing_sign} Real-space pairing-pairing correlation function [see \cref{eq:pairing_corr}] for the $8\times 8$ Hubbard model at $U=8.0$, $t'=-0.2$, and doping ${\delta=1/8}$, obtained with the determinant \textit{ansatz} before [panel (a)] and after [panel (b)] symmetry restoration. Below each panel we show the one-dimensional cut ${C_p(x,y=3)}$.}
    \end{center}
\end{figure}
%%%%%%%%%%%%%%%%%%%%%%%%%%%%%%%%

%%%%%%%%%%%%%%%%%%%%%%%%%%%%%%%%
\begin{figure*}[ht]
    \begin{center}
\centerline{\includegraphics[width=1.8\columnwidth]{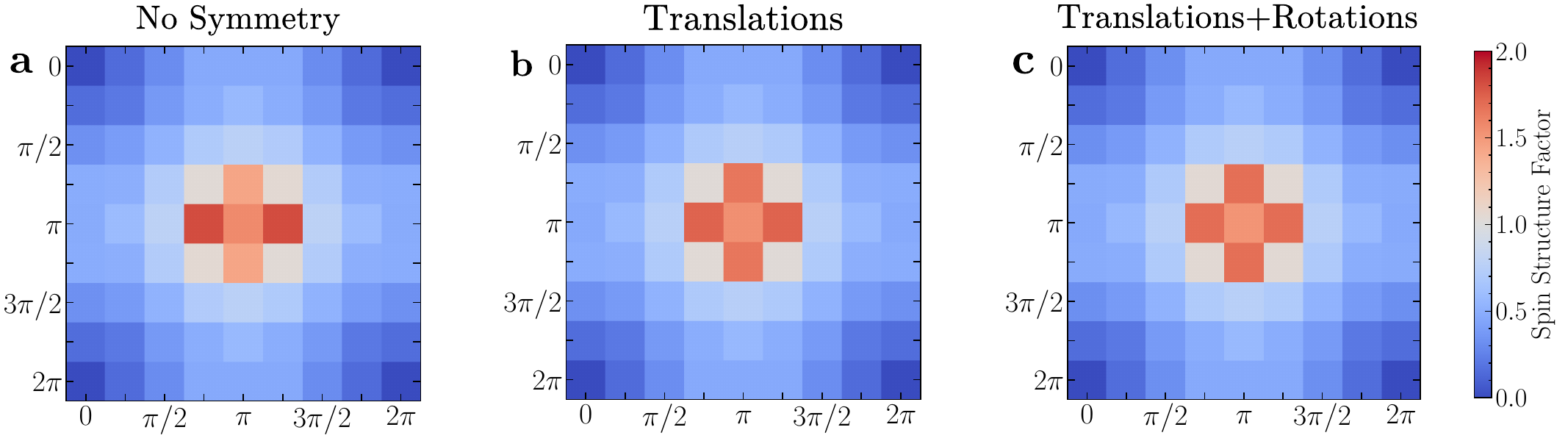}}
        \caption{\label{fig:spin_structure_factor}Spin structure factor for the $8\times 8$ Hubbard model at $U=8.0$, $t'=-0.2$, and doping $\delta=1/8$, obtained with the Pfaffian {\it ansatz} at different levels of symmetry restoration. \textbf{Panel a:} Result without symmetry projection. \textbf{Panel b:} Result after restoring translational symmetry. \textbf{Panel c:} Result after restoring both translational and rotational symmetries.}
    \end{center}
\end{figure*}
%%%%%%%%%%%%%%%%%%%%%%%%%%%%%%%%

\subsection{Pairing Correlation Function}\label{sec:pairing_sign}
In \cref{fig:pairing_sign}, we show the real-space pairing-pairing correlation function defined in \cref{eq:pairing_corr}, obtained with the determinant \textit{ansatz} before and after symmetry restoration. Panel~(a) shows the unsymmetrized state, which displays a pronounced stripe pattern [see \cref{fig:stripe_order}] and only weak superconducting correlations [see \cref{fig:pairing}]. Panel~(b) shows the fully symmetrized state obtained after restoring both translational and rotational symmetries.

Beyond the different long-distance behavior discussed in \cref{sec:pairing}, a key difference emerges in the signs of the pairing correlation function. In the unsymmetrized state, correlations among next-nearest neighbor sites are negative, whereas after symmetry restoration all pairing correlations become positive. This negative short-distance structure appears only before symmetry projection, namely in the regime where stripe order is strongest and superconducting correlations are weakest. We therefore interpret it as a manifestation of the bias of the unsymmetrized determinant state, which favors stripe order at the expense of superconducting correlations. After symmetry restoration, the pairing correlation function becomes consistent, both in sign and in magnitude, with that obtained from the other variational states, namely the particle-hole determinant and the Pfaffian (not shown here).

\subsection{Spin structure factor}\label{sec:spin_structure_factor}
In \cref{fig:spin_structure_factor} we show the spin structure factor, defined as the Fourier transform of the real-space spin-spin correlations ${S(\boldsymbol{k})=\sum_{\boldsymbol{r}} e^{i\boldsymbol{k}\cdot\boldsymbol{r}}
\langle \hat{S}^{z}_{\boldsymbol{0}}\hat{S}^{z}_{\boldsymbol{r}}\rangle}$, which we use to characterize magnetic ordering.
A dominant peak at $\boldsymbol{k}=(\pi,\pi)$ signals antiferromagnetic order. In the present case, however, the structure factor also exhibits four additional incommensurate peaks, which are the fingerprint of stripe order. For the $8\times 8$ cluster considered here, these peaks are located at $\boldsymbol{k}_1=\left(\pi+\frac{\pi}{4},\pi\right)$, $\boldsymbol{k}_2=\left(\pi-\frac{\pi}{4},\pi\right)$,
$\boldsymbol{k}_3=\left(\pi,\pi+\frac{\pi}{4}\right)$ and 
$\boldsymbol{k}_4=\left(\pi,\pi-\frac{\pi}{4}\right)$.
For a fully symmetric wave function, the values of the structure factor at these four momenta are exactly equal by symmetry (see panel (c) of \cref{fig:spin_structure_factor}). This is not necessarily the case for wave functions without symmetry projection, where the four peaks may have different weights (see panel (a) of \cref{fig:spin_structure_factor}). In order to adopt a definition that is valid in all cases and allows for a direct comparison between symmetric and non-symmetric states, we define the stripe magnetic order parameter as the average over the four peak values 
\begin{equation}
M^2_{\text{str}}=\frac{1}{4N}\left[S(\boldsymbol{k}_1)+S(\boldsymbol{k}_2)+S(\boldsymbol{k}_3)+S(\boldsymbol{k}_4)\right] \ .
\end{equation}
For fully symmetric wave functions, it reduces to the common peak value divided by $N$, while it remains well defined also in the absence of symmetry restoration.

\subsection{Bounds on the Order Parameters}\label{sec:bounds}

In \cref{tab:order_params}, we report the $d$-wave superconducting and stripe order parameters obtained from the three fully symmetrized variational states, with both translational and rotational symmetries restored.
%%%%%%%%%%%%%%%%%%%%%%%%%%%%%%%%%%%%%%%%%%%%%
\begin{table}[t]
\centering
\setlength{\tabcolsep}{14pt}
\begin{tabular}{| c | c | c |}
\hline\hline
\textbf{Ansatz} & \textbf{$d$-wave} $\Delta_{\text{SC}}$ & \textbf{Stripe} $M^2_{\text{str}}$ \\
\hline
Det      & 0.0177(1) & 0.0272(1)  \\
\hline
{Pfaffian} & {0.0225(2)} & {0.0263(1)} \\
\hline
Det-PH   & 0.0254(2) & 0.0247(1) \\
\hline\hline
\end{tabular}
\caption{Stripe ($M^2_{\text{str}}$) and $d$-wave superconducting ($\Delta_{\mathrm{SC}}$) order parameters for the $8\times 8$ Hubbard model with PBC at $U=8.0$, $t'=-0.2$, and $\delta=1/8$, after restoring both translational and rotational symmetries. Values are readings from \cref{fig:order_par}. The definitions of the order parameters are given in the main text.}
\label{tab:order_params}
\end{table}
%%%%%%%%%%%%%%%%%%%%%%%%%%%%%%%%%%%%%%%%%%%%%
For the $d$-wave superconducting order parameter $\Delta_{\mathrm{SC}}$ (see the definition in the main text), the determinant {\it ansatz} systematically underestimates pairing correlations. Even after symmetry restoration, a residual bias remains, so its value is naturally interpreted as a \emph{lower bound} on the exact value. By contrast, the particle-hole determinant naturally favors pairing and tends to overestimate superconducting correlations; its symmetry-restored value can therefore be regarded as an \emph{upper bound}. We therefore expect the exact superconducting order parameter $\Delta_{\text{SC,ex}}$ to lie in the range:
\begin{equation}
0.0177(1) \leq \Delta_{\text{SC,ex}} \leq 0.0254(2) \ .
\end{equation}
The situation is reversed for the stripe order parameter $M^2_{\text{str}}$ (see the definition in the main text). In this case, the determinant {\it ansatz}, which naturally favors magnetic order, provides the \emph{upper bound}, whereas the particle-hole determinant, which suppresses spin correlations in favor of pairing, provides the \emph{lower bound}. Accordingly, the exact stripe order parameter $M^2_{\text{str,ex}}$ is expected to be in the interval:
\begin{equation}
0.0247(1) \leq M^2_{\text{str,ex}} \leq 0.0272(1) \ .
\end{equation}
The narrowness of these intervals indicates that symmetry restoration largely removes the initial architectural biases of the different variational states.

\subsection{Optimization Setup}
The variational states are optimized within the Variational Monte Carlo framework~\cite{becca2017} using Stochastic Reconfiguration (SR)~\cite{Sorella1998} with the linear algebra trick~\cite{rende2024stochastic,chen2024empowering}. 
The optimization is carried out for $4\times 10^4$ steps without the symmetries, $2\times 10^3$ steps when restoring the translational symmetry, and an additional $10^2$ steps when restoring both translational and rotational symmetry. We use $M=2^{13}$ Monte Carlo samples per iteration. The learning rate is initialized to $\eta=0.03$ and is gradually annealed during the optimization.

\begin{acknowledgments}
We thank Shiwei Zhang, Markus Holzmann and Ao Chen for useful discussions.
The Flatiron Institute is a division of the Simons Foundation. This work was supported as part of the ``Swiss AI initiative'' by a grant from the Swiss National Supercomputing Centre (CSCS) under project ID a117 on Alps. LLV is supported by SEFRI under Grant No. MB22.00051 (NEQS - Neural Quantum).
\end{acknowledgments}

\bibliography{ref}

\clearpage

%\bibliography{ref}

\end{document}